\documentclass[ aps,%
12pt,%
floatfix,%
final,%
notitlepage,%
oneside,%
onecolumn,%
nobibnotes,%
nofootinbib,%
superscriptaddress,%
showpacs,%
centertags]%
{revtex4}

\begin{document}

\author{A.V. Berezhnoy}\email{aber@ttk.ru}
\affiliation{Moscow State University}
\author{S.N Koshkarev}\email{koshkarev.sergey@gmail.com}
\author{A.V. Luchinsky}\email{Alexey.Luchinsky@ihep.ru}
\author{V.I. Nikolaenko}
\affiliation{Institute for High Energy Physics, Protvino, Russia}

\title{Lepton pair production in muon scattering by nucleus}

\begin{abstract}
Coherent production of lepton pair in muon scattering by nucleus is considered. With the help of different approaches (equivalent photon approximation, direct numerical and analytical calculations) we obtain total cross section of this reaction and distributions over invariant masses of final leptons. Special attention is paid to influence of final lepton identity on total and differential cross sections. We also study the role of coherence  condition. Total cross sections of charmonia production in muon interaction with electromagnetic field of nucleus are also presented.
\end{abstract}

\maketitle

\section{Introduction}

In our article we consider total and differential cross sections of lepton pair production in muon scattering  by nucleus
\begin{eqnarray}
  \mu^+\N &\to& \mu^+\N\mu^+\mu^-.
  \label{reac:muN}
\end{eqnarray}
This subject is discovered thoroughly already (see, for example, \cite{Kelner:1967aa,Kelner:1998mh,Kelner:2000va,Burkhardt:2002vg}), but there are some questions, that deserve more detailed investigation.

One of this questions is the influence of final lepton identity on total and differential cross sections. In mentioned above works it was shown, that total cross sections are changed slightly when this effect is taken into account. Since final lepton identity strongly complicates calculations (for example, the number of Feynman diagrams is doubled), in the following it was often neglected. Actually, the following process was considered
\begin{eqnarray*}
    \mu^+\N &\to& \mu^+\N\ell^+\ell^-
\end{eqnarray*}
with mass of $\ell$-lepton equal to that of muon. In our article we will show, that, though final lepton identity does not change total cross section significantly, invariant mass distributions are changed dramatically. This difference is important for calculation of electromagnetic background to charmonia production in muon scattering by nucleus.

The second question is the influence of coherent condition (that is the  requirement that the nucleus stays undestroyed after muon scattering). This condition can be described by nucleus electromagnetic forfactor, but there is not enough information about its explicit form. In the framework of equivalent photon approximation (EPA) method, that will be used in our paper, there is a simple way to check this condition. One only needs to take a suitable distribution function of equivalent photons.

In second and third sections of our article we calculate total and differential cross sections of the process \eqref{reac:muN} using three different approaches --- EPA, analytic expression, given in \cite{Budnev:1974de}, and direct  calculation. In fourth section the influence of coherence condition is studied. Finally we consider charmonia production in muon interaction with electromagnetic field of the nucleus.

\section{Total cross section}

\begin{figure}
\psfrag{a}{(a)}
\psfrag{b}{(b)}
\psfrag{c}{(c)}
\includegraphics{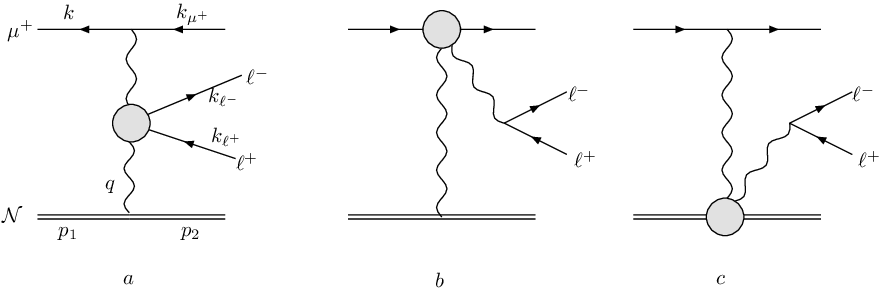}
\caption{Leading order diagrams for $\mu\N\to\mu\N\ell\ell$\label{diag}}
\end{figure}

\begin{figure}
\includegraphics{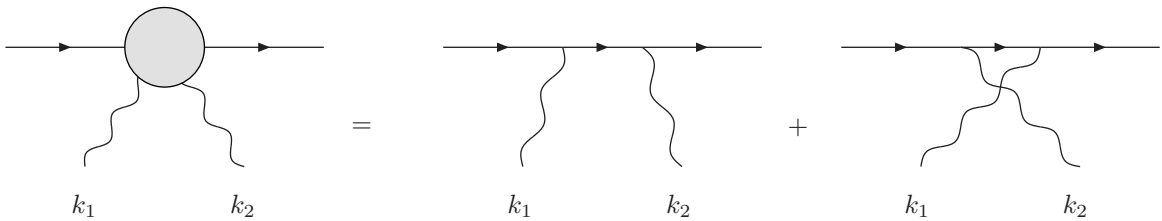}
\caption{$\ell\ell\gamma\gamma$ vertex\label{blob}}
\end{figure}

Leading order diagrams for process
\begin{eqnarray}
  \mu^+(k) \N(p) &\to& \mu^+(k_{\mu^+}) \N(p') \ell^+(k_{\ell^+}) \ell^-(k_{\ell^-})
  \label{reac:mN}
\end{eqnarray}
are shown in figures \ref{diag}, \ref{blob}. If there are identical final particles (i.e. $\ell=\mu$) one should also add diagrams with permutation of
$\ell^+$ and $\mu^+$ lines.

First of all, it is clear that diagram fig. \ref{diag}c does not give significant contribution to the cross section, since corresponding amplitude is
suppressed by nucleus mass $\MN$. It is also clear, that amplitudes of other diagrams are large at small virtuality of the photon emitted by nucleus.
So when calculating the cross section of this reaction one can use Equivalent Photon Approximation (EPA)
\cite{Baur:2001jj,Budnev:1974de,Landau:1934zj}. In the framework of this method the cross section of the process \eqref{reac:mN} is expressed through
the cross section of lepton pair production in muon-photon interaction:
\begin{eqnarray}
  \mu^+(k)\gamma(q) &\to& \mu^+(k_{\mu^+})\ell^+(k_{\ell^+}) \ell^-(k_{\ell^-}).
  \label{reac:muGamma}
\end{eqnarray}
The diagrams describing this process can be obtained from diagrams fig. \ref{diag}a,b after removing nucleus current.

According to equivalent photon approximation cross sections of reactions \eqref{reac:mN} and \eqref{reac:muGamma} are connected by the relation
\begin{eqnarray}
  \sigma_{\mu\N} &=& \int \sigma_{\mu\gamma}(\tm) n(\tm) d\tm,
  \label{eq:EPA}
\end{eqnarray}
where $\tm$ is the invariant mass of final leptons ($\tm^2=(k_{\ell^-}+k_{\mu^+}+k_{\ell^+})^2$) and $n(\tm)$ is the distribution function of equivalent photons in
nucleus. In logarithmic approximation this function is\footnote{
The difference between this expression and function given in \cite{Landau:1934zj} is explained by the variable transformation. We use the invariant
mass $\tm$ instead of virtual photon energy
}
\begin{eqnarray}
  n(\tm) d\tm &=& \frac{4Z^2\alpha}{\pi} \ln\left(
    \frac{m}{\MN}\frac{s}{\tm^2}
  \right)\frac{d\tm}{\tm},
  \label{eq:n}
\end{eqnarray}
where $\alpha$ is fine structure constant, $Z$ is nucleus charge number, $s=(k+p)^2$, $m$ and $\MN$ are muon and nucleus masses respectivelly. The
integration domain in  \eqref{eq:EPA} is determined from $\mu\gamma\to3\mu$ threshold and the condition of positivity of
distribution \eqref{eq:n}:
\begin{eqnarray*}
  (3m)^2 &<& \tm^2< \frac{m}{M}s.
\end{eqnarray*}

\begin{figure}
\psfrag{x}{$\tm$, GeV} \psfrag{y}{$\sigma_{\mu\gamma}(\tm)$, nb}
\includegraphics{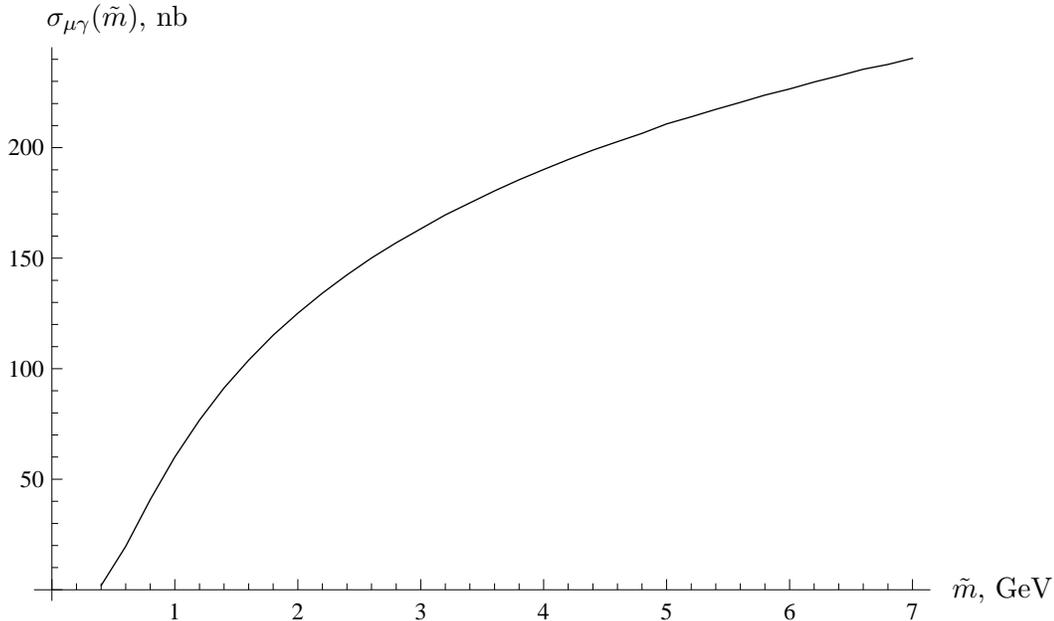}
\caption{Total cross section of $\mu\gamma\to3\mu$ reaction as a function of $\tm$\label{fig:hSigma}}
\end{figure}

In fig. \ref{fig:hSigma} we show $\mu\gamma\to3\mu$ cross section calculated with numerical integration\footnote{
In our calculations we have used two methods: COMPHEP \cite{Boos:2004kh} and method described in \cite{Berezhnoy:2006mz}. The difference between them
is that in the first case squared matrix element is computed analytically, while in the second method matrix element computation is done numerically
and the resulting value is squared. It is clear, that for large number of diagrams the latter approach requires much less numerical calculations and,
hence, leads to smaller errors. The results of these methods coincide.%
}
as a function of $\tm$.  Using this cross section and formulae \eqref{eq:EPA}, \eqref{eq:n} we have computed total cross section of muon pair
production in muon scattering by lithium (that is the nucleus with charge and mass numbers equal to $Z=3$ and $A=7$). It is interesting to mention,
that for fixed muon energy in laboratory frame (that is the rest frame of initial nucleus) this cross section does not depend on nucleus mass. So it is
convenient to introduce a dimensionless variable
\begin{eqnarray*}
  x &=& \frac{s}{m \MN}=2\gamma,
\end{eqnarray*}
where $\gamma$ is the lorentz factor of initial muon in laboratory frame.

This cross section should be compared with the results of analytic calculation of $\mu\N\to\mu\N\mu\mu$ cross section. In work \cite{Budnev:1974de} the
reaction $ep\to epee$ was considered. In the logarithmic approximation the cross section of this reaction is
\begin{eqnarray}
\sigma(ep\to epee) &=& \frac{28}{27\pi}\frac{\alpha^4}{m_e^2}\left(
  L^3-A L^2+B L+C
\right),
\label{eq:Log3}
\end{eqnarray}
where
\begin{eqnarray}
  L &=& \ln\left(\frac{s}{m_e M_p}\right),
\end{eqnarray}
and coefficients $A$, $B$, $C$ are equal to
\begin{eqnarray*}
  A  &\approx& 6.36,\quad   B  \approx 2.6,\quad  C  \approx 40.
\end{eqnarray*}
For large $s$ these coefficients do not depend strongly on particle masses, so for our case we should only change lepton and nucleus masses in the above expressions. We would like to mention, that, analogously to EPA, for fixed muon energy this expression does not depend on nucleus mass.

Another way of calculation of $\mu\N\to\mu\N \mu\mu$ cross section is direct numerical calculation. In this case we have observed the dependence of
cross section on nucleus mass. If $\MN$ and $s$ are increased (while muon $\gamma$-factor is kept constant) total cross section of \eqref{reac:mN}
decreases. This decrease is caused by power corrections ($\sim\MN^2/s$) that are neglected in EPA and \eqref{eq:Log3}.

\begin{figure}
\psfrag{x}{$x$} \psfrag{y}{$\sigma_{\mu\N}$, nb}
\includegraphics{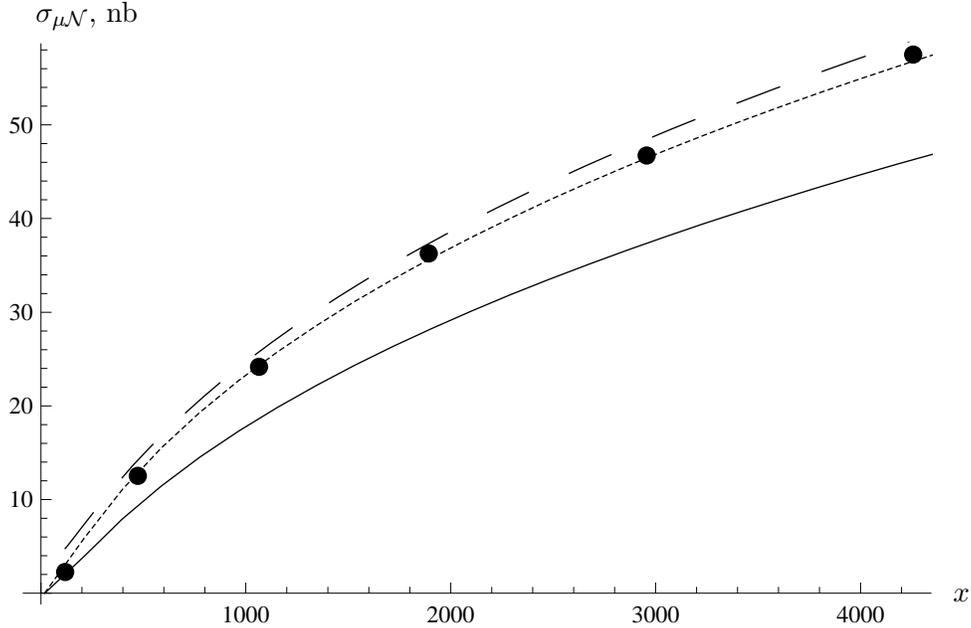}
\caption{ Cross section of $\mu\Li\to3\mu\Li$ reaction. Solid and  dotted lines are EPA results with distribution functions \eqref{eq:n} and corrected
distribution function (see text for details); dashed line is the result of formula \eqref{eq:Log3}, dots --- direct calculation. \label{fig:Sigma}}
\end{figure}

In fig. \ref{fig:Sigma} we show $\mu\N\to3\mu\N$ total cross section as a function of $x$. It can be seen, that the results of direct calculation
(dots) and formula \eqref{eq:Log3} (dashed line) are close to each other (the difference for low $x$ is explained by power corrections), while EPA
cross section (solid line) is smaller. This discrepancy however, can be removed if we multiply the argument of the logarithm in \eqref{eq:n} by $\sim
1.4$ (dotted line in fig. \ref{fig:Sigma}). Since formula \eqref{eq:n} was obtained in logarithmic approximation, this argument is defined up to order
one factor, so such transformation is allowed.

Let us discuss now the influence of final lepton identity on total cross sections of reactions \eqref{reac:mN} and \eqref{reac:muGamma}. The amplitude
of the process $\mu^+\gamma\to\mu^+\mu^-\mu^+$ is obtained from diagrams shown in fig.\ref{diag}a,b (after removing nucleus current and
anti-symmetrizing identical final lepton legs) and can be written as
\begin{eqnarray*}
  \tAT(\mu,\ell) &=& \tANT(\mu,\ell) - \tANT(\ell,\mu),
\end{eqnarray*}
where $\tANT$ is the amplitude of process \eqref{reac:muGamma} without final lepton identity, and symbols $\mu$ and $\ell$ designate momenta and
helicity variables of $\mu^+$- and $\ell^+$-leptons respectively. Total cross section is equal to
\begin{eqnarray}
   \tsigmaT
   &=& \frac{1}{2}\int d\Phi|\tAT|^2 =
    \frac{1}{2}\int d\Phi\left\{     |\tANT(\mu,\ell)|^2 + |\tANT(\ell,\mu)|^2   \right\} -
\nonumber \\ & - &
   \frac{1}{2}\int d\Phi\left\{
     \tANT(\mu,\ell){\tANT}^*(\ell,\mu) + {\tANT}^*(\mu,\ell)\tANT(\ell,\mu)
   \right\},
   \label{eq:sigma1}
 \end{eqnarray}
 where integration is performed over the loretz-invariant phase space of final particles (averaging over their polarizations is also implied), and
factor 1/2 is caused by identical leptons in final state. Since integration domain is symmetric under permutation $\mu^+\leftrightarrow\ell^+$,
expression \eqref{eq:sigma1} can be rewritten as
\begin{eqnarray}
  \tsigmaT &=& \int d\Phi |\tANT(\mu,\ell)|^2 -
    {\mathrm Re} \int d\Phi\left\{\tANT(\mu,\ell) {\tANT}^*(\ell,\mu)\right\} =
\nonumber \\ &=&
  \tsigmaNT + \delta\sigma^{\mu\gamma}.
  \label{eq:sigma2}
\end{eqnarray}
First term here is the cross section with final lepton identity neglected, and second term describes the corrections caused by this effect. It is easy
to understand, that for large $s$ this correction is small. In the region of low invariant mass $m_{\ell^+\ell^-}=\sqrt{(k_{\ell^+}+k_{\ell^-})^2}$ the virtuality of
show in fig. \ref{diag}b photon is small and the amplitude $\tANT(\ell,\mu)$ is enhanced. So in this region both factors in the first term of
expression \eqref{eq:sigma2} are large. This is not the fact for interference contribution (second term in \eqref{eq:sigma2}) since there is no
enhancement for $\tANT(\ell,\mu)$ in this region.

We would like to stress, that presented above arguments do not depend on the dynamics of specific process. The only requirements is symmetry of phase
space integration domain under permutation of final particles and an enhancement of the amplitude on the border of this domain. Near the threshold the
interference term will be comparable with $\tsigmaT$ and final lepton identity will play significant role. This is because for low $\tm$ total
integration domain is comparable with the width of the region, where amplitude $\tANT$ is enhanced.

\begin{figure}
\psfrag{x}{$\tm$, GeV}
\includegraphics{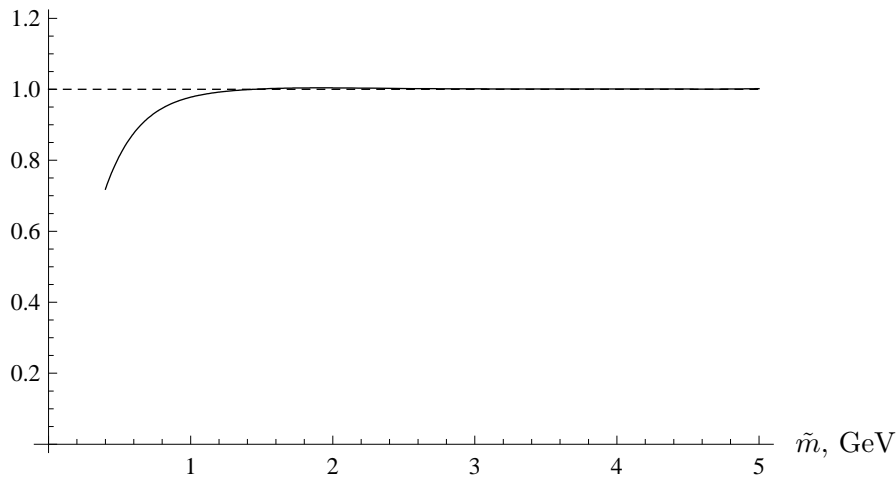}
\caption{
The ratio of $\mu\gamma\to 3\mu$ cross sections, calculated with and without final lepton identity \label{fig:ratio}}
\end{figure}

In figure \ref{fig:ratio} we show the ratio of $\mu\gamma\to3\mu$ cross section with final lepton identity taken into account and neglected. This
figure agrees well with out expectations. Since in EPA cross sections of reactions  $\mu\gamma\to 3\mu$ and $\mu\N\to 3\mu\N$ are related by formula
\eqref{eq:EPA}, for the latter process the behavior of such ratio will be the same.

\section{Invariant massdistributions}

Let us now discuss distributions over invariant masses
\begin{eqnarray*}
  \tm^2 &=& (k_{\ell^-}+k_{\mu^+}+k_{\ell^+})^2, \\
  m_{\mu^+\ell^+}^2 &=& (k_{\mu^+}+k_{\ell^+})^2, \\
  m_{\mu 1}^2 &=& (k_{\mu^+}+k_{\ell^-})^2, \\
  m_{\ell 1}^2 &=& (k_{\ell^+}+k_{\ell^-})^2.
\end{eqnarray*}
In the case of identical leptons last two variables coincide.

\begin{figure}[b]
\psfrag{x}{$\tm$, GeV}
\psfrag{y}{\hskip -1cm $\frac{d\sigma_{\mu\N}}{d\tm}, \frac{\mathrm{nb}}{\mathrm{GeV}}$}
\includegraphics{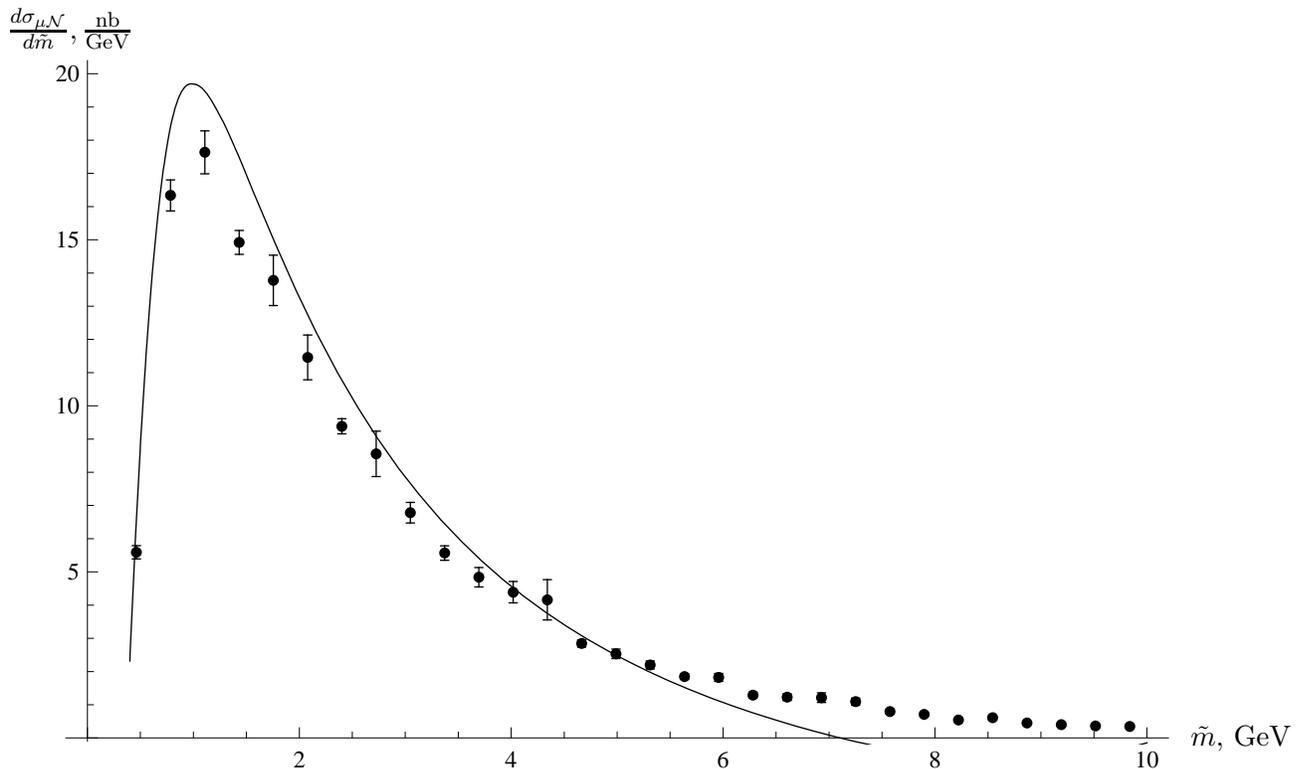}
\caption{Distribution of $\mu\Li\to\mu\Li\mu\mu$ cross section over invariant mass of three final leptons \label{fig:dSigmadM3}}
\end{figure}

In the following we will consider the scattering of 160 GeV muon by lithium. The distribution over mass $\tm$ follows directly from eq.
\eqref{eq:EPA}. In figure \ref{fig:dSigmadM3} we show this distribution calculated with EPA (soled line) and direct calculation (dots). From previous section (see figure \ref{fig:ratio}) it is clear, that final lepton identity has little effect on this distribution.

\begin{figure}
\psfrag{a}{(a)} \psfrag{c}{\hskip -1cm$\frac{d\sigma_{\mu\N}}{dm_{\mu\ell}}$, $\frac{\mathrm{nb}}{\mathrm{GeV}}$}
\psfrag{b}{(b)} \psfrag{d}{I/NI}
\includegraphics{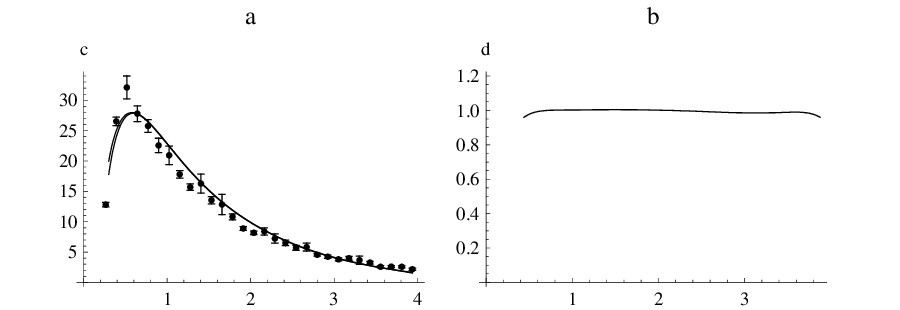 }
\caption{
(a) --- distribution of $\mu\Li\to\mu\Li\mu\mu$ cross section over $m_{\mu^+\ell^+}$; (b) --- the ratio of distributions with final lepton identity
neglected and taken into account
\label{fig:dSigmadMpp}}
\end{figure}

Distribution of muon-nucleus cross section over invariant mass of $\mu^+\ell^+$-pair in EPA is calculated with formula similar to \eqref{eq:EPA}:
\begin{eqnarray*}
  \frac{d\sigma^{\mu\N}}{dm_{\mu\ell}} &=& \int \frac{d\sigma^{\mu\gamma}}{dm_{\mu\ell}} n(\tm) d\tm.
\end{eqnarray*}
This distribution and the ratio of differential cross sections with and without final lepton identity are show in figure \ref{fig:dSigmadMpp}. It is
clear, that in this case the influence of identity is also small. The reason is clear --- the integration domain is symmetric under the permutation of
$\mu^+$ and $\ell^+$ momenta.

\begin{figure}
\psfrag{x}{$m_{\mu^+\mu^-}$, GeV}
\psfrag{y}{\hskip -.5cm $\frac{d\sigma}{dm_{\mu^+\mu^-}}\,\frac{\mathrm{nb}}{\mathrm{GeV}}$}
\includegraphics{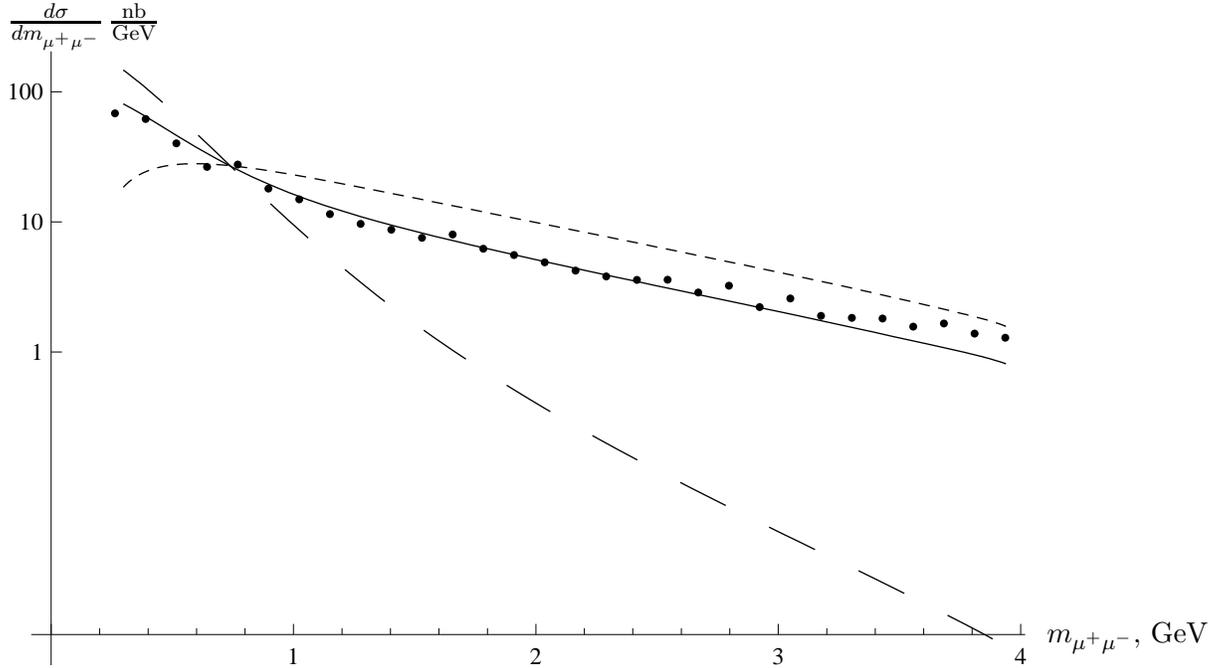}
\caption{
Distributions of $\mu\Li\to\mu\Li\mu\mu$ cross section over invariant mass of leptons with opposite charge. Solid line --- distribution with final lepton identity taken into account; dashed and dotted lines --- distributions over $m_{\mu^+\ell^-}$ and $m_{\ell^+\ell^-}$; dots --- results of direct calculation
\label{fig:MpMm}}
\end{figure}

\begin{figure}
\psfrag{x}{$m_{\mu^+\ell^-}$} \psfrag{y}{$m_{\ell^+\ell^-}$}
\includegraphics{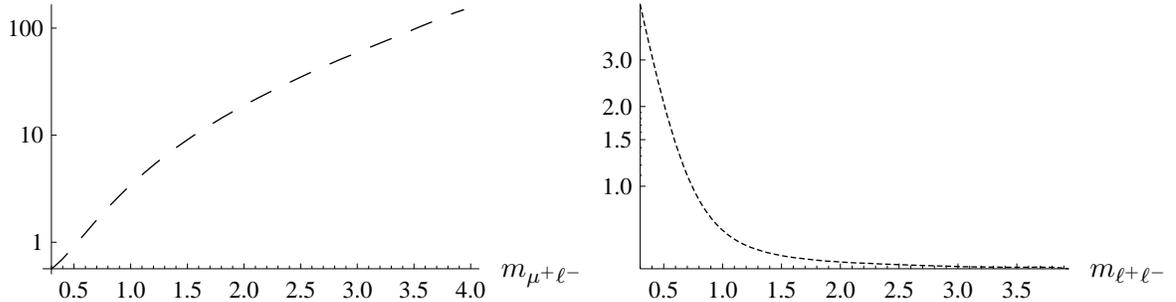}
\caption{Ratio of $\mu\Li\to\mu\Li\mu\mu$ cross section distributions with final lepton identity taken into account and neglected
\label{fig:MpMmRatio}
}
\end{figure}

If we consider the distribution over invariant mass of leptons with opposite charges (i.e. $\mu^+\mu^-$) the situation is completely different. When final lepton identity is neglected, we actually have two different variables --- $m_{\ell^+\ell^-}$ and $m_{\mu^+\ell^-}$. If identity is taken into account, theses variables coincide. In figure \ref{fig:MpMm} we show distributions over these three masses. It can be clearly seen (see fig. \ref{fig:MpMmRatio}, where the ratio of distribution with and without final lepton identity is shown), that final lepton identity changes these distributions dramatically. If we measure, for example, the distribution over $(\mu^+\mu^-)$ pair mass in the region of $M_{J/\psi}$ with an instrumental error  $\Delta=50$ MeV, we get
\begin{eqnarray}
  \left. \Delta \frac{d\sigma^{\mu\N}}{dm_{\mu1}} \right|_{m_{\mu^+\ell^-}=M_{J/\psi}} &=& 1.5\,\mathrm{pb}
  \label{eq:fracNT}
\end{eqnarray}
when final lepton identity is neglected and
\begin{eqnarray}
  \left. \Delta \frac{d\sigma^{\mu\N}}{dm_{\mu1}} \right|_{m_{\mu^+\ell^-}=M_{J/\psi}} &=& 92\,\mathrm{pb}.
  \label{eq:fracT}
\end{eqnarray}
if it is taken into account. This difference is crucial in determining the electromagnetic background to $J/\psi$ production in muon scattering by
nucleus.

\section{Form-factors and coherence condition}

Up to now we treated nucleus as a point-like charged particle. It is clear, that this approximation has little to do with real world, since nucleus can easily be destroyed in muon scattering.

One can go beyond this approximation by introducing electromagnetic form factor of the nucleus. It is well known (see, for example, \cite{Kuraev:2006ym} ), that for proton this form factor is
\begin{eqnarray*}
  F(q^2) &=& \left(1-\frac{q^2}{Q_0^2}\right)^{-2},\qquad Q_0^2 \approx 0.7\,\mathrm{GeV}^2.
\end{eqnarray*}
For nucleus formfactor is unknown, but we can describe it qualitatively by imposing a restriction on a squared transferred momentum $-q^2<Q_0^2$. In our calculations we have used values $Q_0^2=0.1\,\mathrm{GeV}^2$ and $Q_0^2=0.05\,\mathrm{GeV}^2$.

In the framework of equivalent photon approximation it is more convenient to use another approach \cite{Baur:2001jj}. The dependence on impact parameter $b$ is introduced in distribution function $n$:
\begin{eqnarray*}
  n_K(\omega,b) &=& \frac{Z^2\alpha}{\pi^2} \left\{
    K_1^2\left(\frac{\omega b}{\gamma \beta}\right)+\frac{1}{\gamma^2}K_0^2\left(\frac{\omega b}{\gamma \beta}\right)
  \right\},
\end{eqnarray*}
where $\omega$ is photon energy and $K_{0,1}$ are modified Bessel functions. For nucleus to remain undestroyed the following condition should should be valid: $b>R_\mathrm{min}$. Here $R_\mathrm{min}\approx 1.1 A^{1/3}$ fm is a typical radius of the nucleus. After integration over impact parameter with this restriction we get
\begin{eqnarray}
  n_K(\omega) &=& \frac{2Z^2\alpha}{\pi\omega}\left[
    \xi K_0(\xi) K_1(\xi) - \frac{\xi^2}{2}\left( K_1^2(\xi)-K_0^2(\xi)\right)
  \right]
\label{eq:nK}
\end{eqnarray}
where $\xi=\omega R_\mathrm{min}/\gamma \beta $.

\begin{figure}
\psfrag{x}{$m_{\mu^+\mu^-}$, GeV}
\psfrag{y}{\hskip -1cm$\frac{d\sigma_{\mu\N}}{dm_{\mu^+\mu^-}}$, $\frac{\mathrm{nb}}{\mathrm{GeV}}$}
\includegraphics{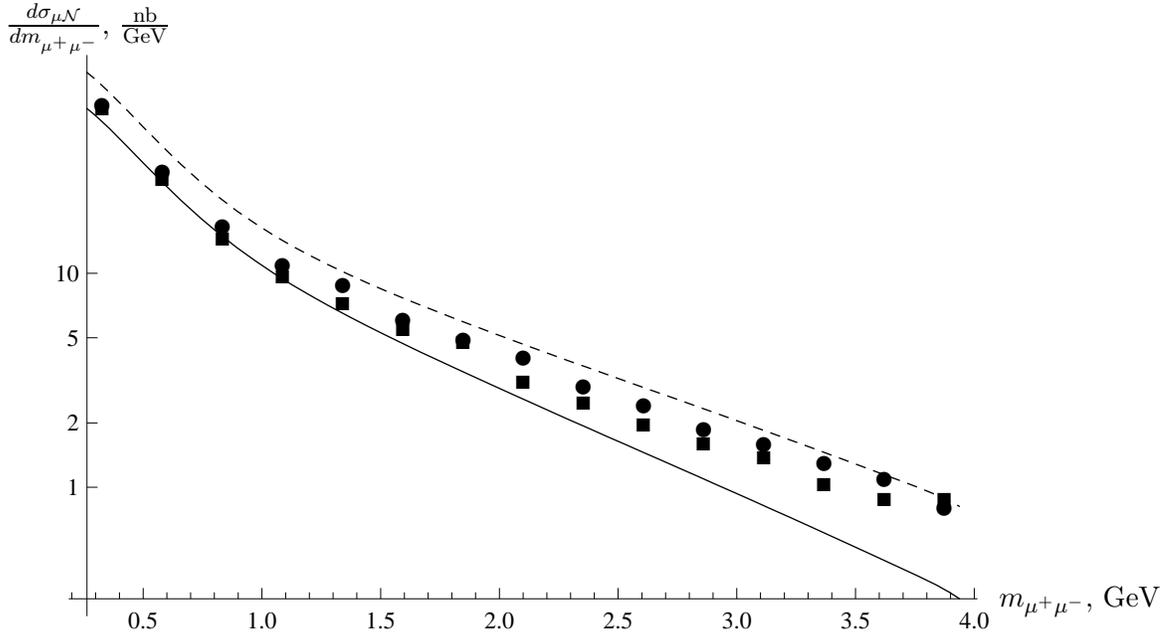}
\caption{
Distribution of $\mu\Li\to\mu\Li\mu\mu$ cross section over invariant mass of $(\mu^+\mu^-)$-pair. Solid line is EPA result with distribution function eqref{eq:nK}, circles and squares are represent direct calculation with $-q^2<0.1\,\mathrm{GeV}^2$ and $-q^2<0.05\,\mathrm{GeV}^2$ respectively, dashed line is EPA in point-like approximation.
\label{fig:SigmaFF}}
\end{figure}

In figure \ref{fig:SigmaFF} we show $m_{\mu^+\mu^-}$ distribution of $\mu\Li\to\mu\Li\mu\mu$ cross section (with final lepton identity taken into account), obtained with EPA (distribution function \eqref{eq:nK} was used) and the results of direct calculations for different values of a cutoff parameter.

\section{$\mu\N \to \mu\chi_c \N$}

\begin{figure}
\psfrag{a}{(a)}\psfrag{b}{(b)}
\includegraphics{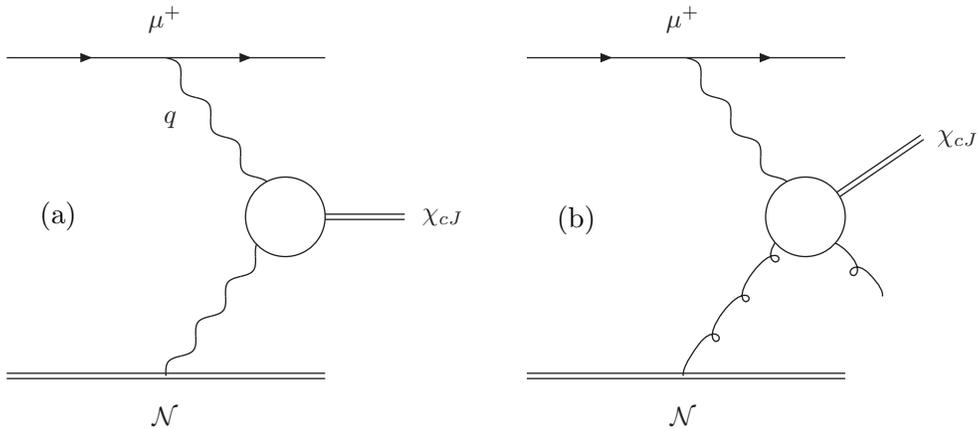}
\caption{
Diagrams of $\chi_c$-meson production in (a) electromagnetic and (b) strong processes.
\label{diagChi}}
\end{figure}

Let us now proceed to charmonium production in muon scattering by nucleus. In particular, we would like to consider $\chi_c$-meson production via the
interaction of muon with equivalent photons (corresponding diagrams are shown in figure \ref{diagChi}a). In comparison with gluon reaction (fig.
\ref{diagChi}b) the cross section of this process is suppressed by electromagnetic constant. In the gluon case, on the other hand, color conservation
requires emission of an additional gluon. It is therefore straightforward to get the following estimate for the ratio of electromagnetic and strong
cross sections:
\begin{eqnarray*}
  \frac{\sigma(\mu\gamma\to\mu\chi_{cJ})}{\sigma(qg\to q\chi_{cJ})} &\sim&
  \frac{\alpha}{\alpha_s^2} \sim 10^{-1}.
\end{eqnarray*}

In paper \cite{Likhoded:2007fz} one can find the expressions for differential cross sections of charmonium production in quark-gluon interaction. It is
evident, that in our case we should remove from these expressions all color factors (including factors caused by color degrees of freedom of initial
particles) and change $\alpha_s$ to $\alpha$. As a result we get
\begin{eqnarray}
  \frac{d\sigma(\mu\gamma\to\mu\chi)}{dq^2} &=&
  2N_c\frac{\alpha^3}{\alpha_s^3}
  \frac{d\sigma(qg\to q\chi)}{dq^2}.
\label{eq:sigm}
\end{eqnarray}
There is another difference between these two reactions. In our case, contrary to scattering on light quark, we cannot neglect muon mass. The equivalent
photon approximation, used used in our paper, give results with logarithmic accuracy, so there is no need to leave muon mass in the differential cross
section \eqref{eq:sigm}. The influence of this mass on the integration domain is, however, crucial. The lower bound on the squared tranfered momentum
is
\begin{eqnarray}
  \hat t_\mathrm{min} &\approx& -\frac{M_\chi^4 m_\mu^2}{\hat s(\hat s-M_\chi^2)}, \label{eq:tmin}
\end{eqnarray}
where $M_\chi$ is $\chi_c$-meson mass (we neglect the difference between $\chi_{c0}$, $\chi_{c1}$ and $\chi_{c2}$ masses). In the limit $m_\mu=0$ we have $\hat t_\mathrm{min}=0$. Cross section \eqref{eq:sigm} diverges logarithmicaly  in this region, so muon mass should be
taken into account.

\begin{figure}
\psfrag{a}{$\sqrt{\hat s}$, GeV}
\psfrag{b}{$\hskip -1.5cm \sigma(\mu\gamma\to\mu\chi_{cJ})$, pb}
\includegraphics{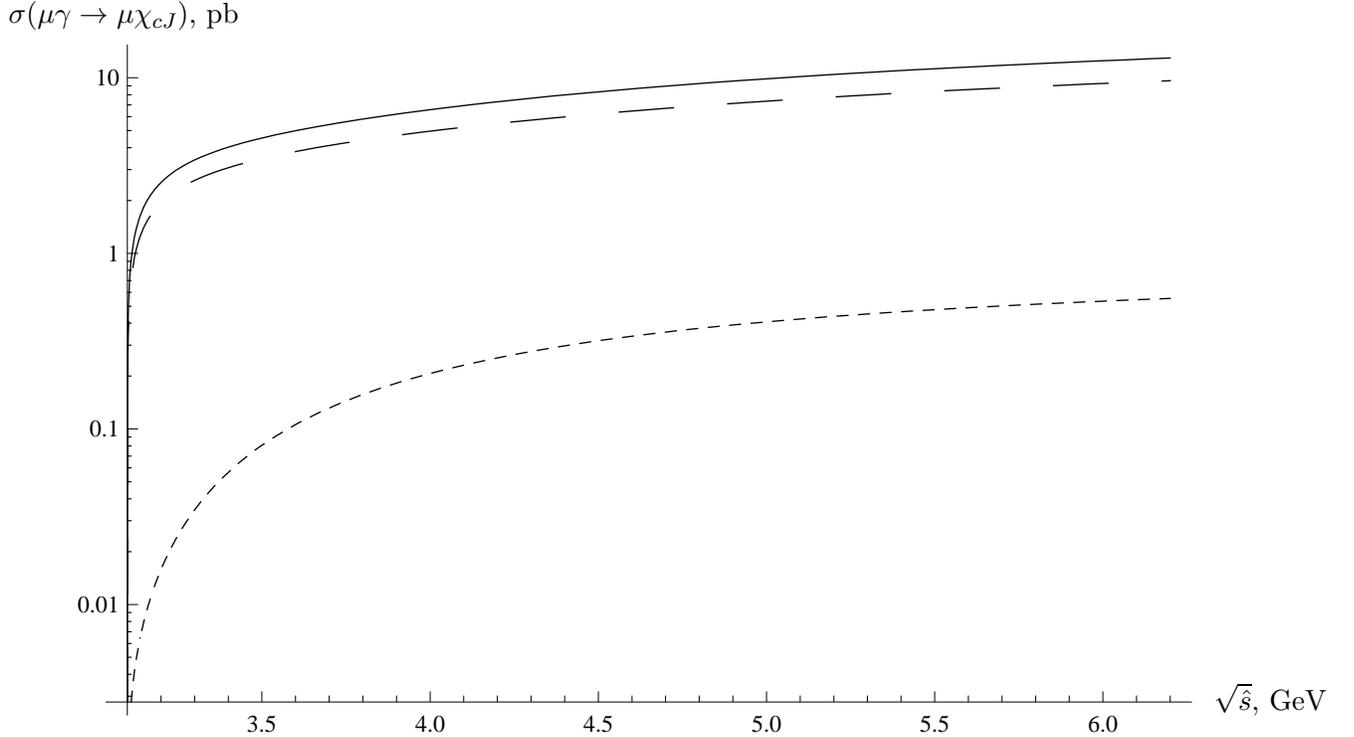}
\caption{Solid line --- $J=2$, dashed line --- $J=0$, dotted line --- $J=1$. \label{fig:chi}}
 \end{figure}

In figure \ref{fig:chi} the dependence of the $\mu\gamma\to\mu\chi$ cross section on $\mu\gamma$ energy is shown. After integrating with distribution
function of equivalent photons we obtain the cross section of the reaction $\mu \Li\to\mu \Li\chi_J$:
\begin{eqnarray*}
  \sigma(\mu \Li\to \mu\, \Li\,\chi_{c0})&=&  14\,\mathrm{fb}, \\
  \sigma(\mu \Li\to \mu\,\Li\,\chi_{c1})&=&  0.5\,\mathrm{fb}, \\
  \sigma(\mu \Li\to \mu\,\Li\,\chi_{c2})&=&  19\,\mathrm{fb}.
\end{eqnarray*}
Produced $\chi$-mesons decay subsequently into experimentally observed lepton pair ($\chi_c\to J/\psi\gamma\to\mu^+\mu^-\gamma$). If we neglect permutation background, than the cross section of
this process is
\begin{eqnarray*}
  \sigma(\mu \Li\to\mu \Li\chi\to\mu \Li\gamma\mu^+\mu^-) &=&  \sigma(\mu \Li\to\mu \Li\chi)\Br(\chi\to\gamma J/\psi)\Br(J/\psi\to\mu^+\mu^-).
\end{eqnarray*}
For different values of $\chi_c$-meson spin we have
\begin{eqnarray*}
  \sigma(\mu \Li\to\mu \Li \chi_{c0}\to\mu \Li \gamma \mu^+\mu^-) &=& 8\cdot  10^{-3}\ \mathrm{fb}, \\
  \sigma(\mu \Li\to\mu \Li \chi_{c1}\to\mu \Li \gamma \mu^+\mu^-) &=& 11\cdot 10^{-3} \ \mathrm{fb}, \\
  \sigma(\mu \Li\to\mu \Li \chi_{c2}\to\mu \Li \gamma \mu^+\mu^-) &=& 0.22              \ \mathrm{fb}.
\end{eqnarray*}
It is clear, that these cross sections are smaller, than the cross section of the non-resonant process \eqref{reac:muN}.

\section{Conclusion}

In our article we consider lepton pair production in muon scattering by nucleus. This process was already studied in details in earlier works, but some questions are still opened.

One of such questions is the influence of final lepton identity on total cross section of the process
\begin{eqnarray*}
  \mu^- \N &\to& \mu^- \N \mu^+ \mu^-.
\end{eqnarray*}
and distributions over invariant mass of $\mu^+\mu^-$-pair. In earlier works it was shown that this identity changes total cross section slightly. We
confirm this result. Invariant mass distributions, on the other hand, change significantly when final lepton identity is taken into account. For
example, if muon with energy 160 GeV is scattered by litium, than in the region $m_{\mu^+\mu^-}\approx M_{J/\psi}$ we have
\begin{eqnarray*}
  \left. \frac{d\sigma_'(\mu \Li \to \mu \Li \mu^+\mu^-)}{dm_{\mu^+\mu^-}}\right|_{m=M_{J/\psi}} &=& 1.8\frac{\mathrm{nb}}{\mathrm{GeV}}
\end{eqnarray*}
for identical leptons and
\begin{eqnarray*}
  \left. \frac{d\sigma_{'}(\mu \Li \to \mu \Li \mu^+\mu^-)}{dm_{\mu^+\mu^-}}\right|_{m=M_{J/\psi}} &=& 0.03\frac{\mathrm{nb}}{\mathrm{GeV}}
\end{eqnarray*}
when this effect is neglected. This difference is significant when estimating background to $J/\psi$-meson production in lepton-nucleus scattering.

In the final section we used Weitzaker-Williams method to calculated the cross section of $\chi_c$-meson production in electromagnetic interaction of
muon with nucleus and subsequent decay $\chi\to\gamma\JP\to\gamma\mu^+\mu^-$. Resulting cross sections are significantly smaller, than the cross
sections of the nonresonant processes. So, one can state that charmonium mesons in lepton-nucleus interaction are produced by strong interaction.

We would like to thank A.K. Likhoded for usefull discussions. This work was supported by Russian foundation for basic research,
grant \# 7-20-00417a. The work of A. Berezhnoy was partially supported by by Dynasty Foundation (project for Young Scientist Support)  and INTAS Fellowship Grant for Young Scientists Nr.~05-112-5117.

\end{document}